\documentclass[trackchanges,twocolumn]{aastex701}
\usepackage{amsmath}
\usepackage{rotating}
\usepackage{afterpage}
\usepackage{soul}
\usepackage{svg}
\usepackage{float}

\begin{document}

\title{Understanding the Energy Input Required for Methane Emission on CWISEP J193518.59–154620.3: A Comprehensive Analysis}

\author[orcid=0009-0001-9981-0171, sname='Smith']{Kayla J. Smith}
\altaffiliation{The University of Arizona}
\affiliation{Department of Planetary Sciences, Lunar and Planetary Laboratory, University of Arizona, Tucson, Arizona}
\email[show]{kaylasmith1@arizona.edu}  

\author[orcid=0000-0002-5251-2943, sname='Marley']{Mark S. Marley} 
\altaffiliation{The University of Arizona}
\affiliation{Department of Planetary Sciences, Lunar and Planetary Laboratory, University of Arizona, Tucson, Arizona}
\email{marksmarley@arizona.edu}

\author[orcid=0000-0003-3071-8358, sname='Koskinen']{Tommi T. Koskinen} 
\altaffiliation{The University of Arizona}
\affiliation{Department of Planetary Sciences, Lunar and Planetary Laboratory, University of Arizona, Tucson, Arizona}
\email{tommi@lpl.arizona.edu}

\author[orcid=0000-0001-5864-9599, sname='Mang']{James Mang}
\altaffiliation{The University of Texas at Austin}
\altaffiliation{NSF Graduate Research Fellow.}
\affiliation{Department of Astronomy, University of Texas at Austin, 2515 Speedway, Austin, TX 78712, USA}
\email{j_mang@utexas.edu}

\begin{abstract}
The Y dwarf WISE 1935 exhibits a thermal inversion in its radiative atmosphere, producing methane emission features in its JWST spectrum, but the physical mechanism responsible for this inversion remains unknown. Using the open-source radiative--convective equilibrium code \texttt{PICASO}, we model atmospheric heating with Chapman energy deposition profiles to reproduce the observed thermal inversion and methane emission feature. Our models require heating rates of approximately $10^{5}$--$10^{6}\ \mathrm{erg\ cm^{-2}\ s^{-1}}$. We show that the atmospheric response depends primarily on the integrated heating deposited in the observable atmosphere, revealing a degeneracy between heating magnitude, vertical extent, and emitting surface fraction. Disequilibrium chemistry lowers the required energy input by lowering CH$_4$ opacity and strengthening the inversion. Comparison with recent electron-beam heating models indicates that reproducing the thermal inversion in W1935 requires substantially greater energy deposition than currently predicted for brown dwarf auroral heating, while the observed methane emission favors energy deposition near $10^{-3}$--$10^{-2}$ bar. Our models also predict a prominent methane emission feature near 7.8 $\mu$m, along with energy-sensitive ammonia features near 6 $\mu$m, implying a bolometric luminosity greater than that yet measured. Finally, we investigate potential sources of the inferred upper-atmospheric heating. We find that Joule heating would require a strong magnetic field and large electron densities, the latter supported by external ionization from an unidentified source. We also consider cometary impacts as a possible source of atmospheric heating.

\end{abstract}



\section{Introduction}
\label{sec:1}

Thermal inversions, regions where atmospheric temperature increases with altitude, are a key diagnostic of the energy balance and radiative structure of planetary atmospheres. In the Solar System, stratospheric temperature inversions generally arise when atmospheric absorbers intercept incident solar radiation and deposit energy at high altitudes. On Earth, ozone absorbs ultraviolet (UV) radiation, heating the stratosphere and producing a thermal inversion. Similarly, in the giant planet atmospheres, CH$_4$ and its photochemical products absorb solar radiation, generating analogous stratospheric inversions. Beyond the Solar System, temperature inversions have also been attributed to strong optical absorbers such as TiO in highly irradiated exoplanets \citep[e.g.,][]{2008ApJ...678.1419F,2009ApJ...699.1487S}, while auroral processes driven by energetic particle precipitation \citep[e.g,][]{2000RvGeo..38..295B,Badman2015} provide an additional mechanism for upper atmospheric heating in the giant planets. Understanding the physical processes that generate and sustain temperature inversions is therefore central to modeling the atmospheric structure, chemistry, and observable spectra of planets and planet-like objects more broadly and testing that understanding requires probing whether, and how, inversions can arise in atmospheres that are not irradiated at all.

Those brown dwarfs that do not orbit a host star lack the external irradiation or stellar-driven magnetospheric electron precipitation that can result in atmospheric inversions in irradiated planets and close-in exoplanets. Consistent with this, methane is generally observed in absorption rather than emission in brown dwarf spectra, reflecting a temperature profile that decreases monotonically with altitude as expected for objects radiating only their own internal heat. The cold, free-floating brown dwarfs WISE J1206+8401 \citep{Schneider_2015, Fontantive_2023} and WISE 0855$-$0714 \citep{Luhman_2023, Zapatero_Osorio_2016}, for example, both display strong methane absorption at 3.325~$\mu$m, consistent with this expectation. Among brown dwarfs, Y dwarfs represent the coldest population, with effective temperatures of roughly 250--500~K \citep[e.g.,][]{Helling_2014}, and therefore the atmospheric regime in which internally generated heating mechanisms, if present, should be most detectable.

CWISEP J193518.59$-$154620.3 (WISE 1935 or W1935) is a Y dwarf. First discovered by the Wide-field Infrared Survey Explorer \citep[WISE,][]{2019ApJ...881...17M} and later observed at higher signal-to-noise with \textit{JWST} \citep{Faherty_2024}, W1935 was found via spectral retrieval to have an effective temperature of 482~K, a radius of 0.96~$R_{\mathrm{Jup}}$, $\log g = 4.7 \pm 0.5$, a mass of 6--35~$M_{\mathrm{Jup}}$, a C/O ratio of 0.77, and [M/H] of 0.45 \citep{Faherty_2024}. Critically, its \textit{JWST} spectrum shows methane in emission at 3.325~$\mu$m, the opposite of the absorption seen in WISE J1206+8401 and WISE 0855$-$0714, implying a thermal inversion of $\sim200$~K near 10~mbar in its temperature--pressure profile, subsequently modeled by \citet{Faherty_2024}. W1935 also hosts a confirmed companion at a separation of 2.48~au \citep[]{defurio2025discoverysecondyydwarf}, itself similarly cold, making this the second Y+Y brown dwarf binary discovered by \textit{JWST}, following W0336 \citep[]{Calissendorff_2023}. Because neither member of this binary provides meaningful external irradiation to the other, W1935's inversion cannot be explained by any of the mechanisms that typically produce inversions in irradiated atmospheres, making it a rare, largely irradiation-free laboratory for isolating the process that is actually responsible.

Furthermore, if an isolated, unirradiated brown dwarf can sustain a substantial thermal inversion, then some internal or environmental energy source must be depositing heat into the atmosphere in a way that is not currently captured by standard brown dwarf atmospheric models. Because these models share much of the same underlying physics as those used to interpret directly imaged and self-luminous giant exoplanets, understanding W1935 has implications well beyond a single object. Constraining the energy budget and heating mechanism responsible for its inversion provides an opportunity to test our understanding of atmospheric energy transport and may reveal physical processes that are important in other giant-planet-like atmospheres where non-stellar heating cannot be ignored.

This work aims to constrain the energy required to reproduce the observed temperature inversion in W1935 and its associated spectra presented by \citet{Faherty_2024}, as a critical first step toward identifying the physical mechanism responsible for heating this atmosphere. Section~\ref{sec:2} describes the methodology, including a detailed overview of the modeling framework. Section~\ref{sec:3} presents the results, highlighting the best-fit spectra and T--P profiles, comparing the energy requirements of chemical equilibrium and disequilibrium scenarios, and presenting the predicted spectrum for W1935 along with a fractional emission area analysis. Section~\ref{sec:4} discusses two potential heating mechanisms: Joule heating and cometary impacts. Section~\ref{sec:5} discusses the implications of our results, while Section~\ref{sec:6} summarizes the primary findings and presents our conclusions.

\section{Methods}
\label{sec:2}
We use \texttt{PICASO} (A Planetary Intensity Code for Atmospheric Spectroscopy Observations) to model the atmospheric temperature profile and to constrain the amount of energy required to reproduce the temperature inversion and methane emission feature of W1935 \citep[]{Mukherjee_2023, 2026ApJ..1000...98M}. \texttt{PICASO} is an open source model that can generate 1D radiative--convective-equilibrium (RCE) climate models and synthetic spectra for brown dwarfs and exoplanets. It includes the ability to model atmospheres in both chemical equilibrium and disequilibrium. One of the additional functionalities in the climate model \citep[as originally developed in][]{MARLEY1999268} is the energy injection scheme, in which a localized atmospheric energy source can be injected into the atmosphere, effectively increasing the energy flux that the model atmosphere must transport to space from at and above the energy source. The atmosphere responds to this perturbation by adjusting the temperature profile self-consistently to maintain radiative--convective flux conservation.

\subsection{Chapman Energy Deposition Profile}
\label{sec:2.1}
In this work, we model atmospheric heating using a Chapman energy deposition profile, which provides a physically motivated description of vertically distributed heating from attenuated incident energy \citep{1931PPS....43...26C}. The Chapman prescription is parameterized by the total injected energy, the pressure of peak energy deposition, and the vertical extent of the perturbation. This approach has previously been used to approximate heating from additional opacity sources that absorb incident flux \citep{1931PPS....43...26C, Morley2014patchy}. For Uranus, for example, \cite{MARLEY1999268} found that an energy deposition rate of $10\ \rm erg\  cm^{-2}\ s^{-1}$ centered near 0.01 bar and distributed over approximately one scale height was required to reproduce the observed Voyager radio occultation temperature profile. We aim to make a similar determination here for W1935.

To explore the sensitivity of the atmospheric structure and emergent spectra to atmospheric heating, we construct a grid of Chapman energy deposition models spanning approximately an order of magnitude around the best-fitting solutions. By varying both the magnitude and vertical distribution of the injected energy, we identify the range of models capable of reproducing the observed methane emission feature and thermal inversion in W1935. For all models presented in this work, we fix the brown dwarf's internal heat flux, corresponding to an effective temperature of $482~\mathrm{K}$, and surface gravity of $501~\mathrm{m~s^{-2}}$, consistent with the properties inferred by \citet{Faherty_2024}.

A nominal Chapman profile distributes energy over the local atmospheric pressure scale height, $H$. Because the physical mechanism responsible for the energy deposition in W1935 is unknown, we allow the characteristic vertical width of the Chapman heating profile, $H_{\rm Chapman}$, to differ from the atmospheric scale height. The volumetric heating rate is parameterized as

\begin{equation}
\begin{split}
Q(P) = Q_0 \exp \Bigg[
&1
+ \frac{H}{H_{\rm Chapman}}
\ln\left(\frac{P}{P_{\rm peak}}\right) \\
&-
\left(\frac{P}{P_{\rm peak}}\right)^{H/H_{\rm Chapman}}
\Bigg]
\end{split}
\label{eq:chapman}
\end{equation}

\noindent where $P_{\rm peak}$ is the pressure at which the heating rate reaches its maximum and $Q_0$ is a normalization constant that sets the total injected energy. The free parameter is the ratio $H/H_{\rm Chapman}$. As evident from Equation~(\ref{eq:chapman}), increasing $H/H_{\rm Chapman}$ causes the heating profile to decline more rapidly away from $P_{\rm peak}$, producing a more narrowly concentrated energy deposition profile. Conversely, smaller values of $H/H_{\rm Chapman}$ distribute the injected energy over a broader vertical region.

We explored ten values of the Chapman width parameter,

\[
\begin{split}
\frac{H}{H_{\rm Chapman}} = \{
&4.5,\ 3.5,\ 2.5,\ 2.0,\ 1.5,\\
&1.0,\ 0.8,\ 0.6,\ 0.4,\ 0.2
\}
\end{split}
\]

\noindent corresponding to heating profiles ranging from very narrow ($H/H_{\rm Chapman}=4.5$) to very broad ($H/H_{\rm Chapman}=0.2$). Although all ten values were included in our parameter survey, much of the subsequent discussion focuses on the representative cases $H/H_{\rm Chapman}=4.5$, 1.0, and 0.2, which span the full range of heating widths explored and capture the primary trends in the atmospheric response.

Furthermore, we explored a range of total energy injection rates, spanning $3.0\times10^{5}$, $4.0\times10^{5}$, $6.0\times10^{5}$, $7.5\times10^{5}$, $9.0\times10^{5}$, $1.2\times10^{6}$, $1.5\times10^{6}$, and $2.0\times10^{6}\ \mathrm{erg\ cm^{-2}\ s^{-1}}$. This range was chosen to systematically investigate the magnitude of atmospheric heating required to reproduce the observed spectrum of W1935 and to constrain the order of magnitude of the energy deposition necessary to generate the methane emission feature. For context, the globally averaged conductive heating flux required to maintain Jupiter's thermosphere is approximately $0.6~\mathrm{erg~cm^{-2}~s^{-1}}$ (equivalent to a total heating power of $\sim40$~TW distributed over Jupiter's surface; \citealt{yelle04}). Although this value refers to Jupiter's thermosphere rather than its stratosphere, it provides a useful benchmark for atmospheric energy deposition. By comparison, the energy fluxes required to reproduce W1935's methane emission feature are $10^{5}$--$10^{6}~\mathrm{erg~cm^{-2}~s^{-1}}$ (depending on where the energy is injected and over how many scale heights), approximately six orders of magnitude larger despite W1935 receiving no external stellar irradiation.

\section{Results}
\label{sec:3}

\begin{figure*}[!htb]
\centering
\includegraphics[scale=0.62]{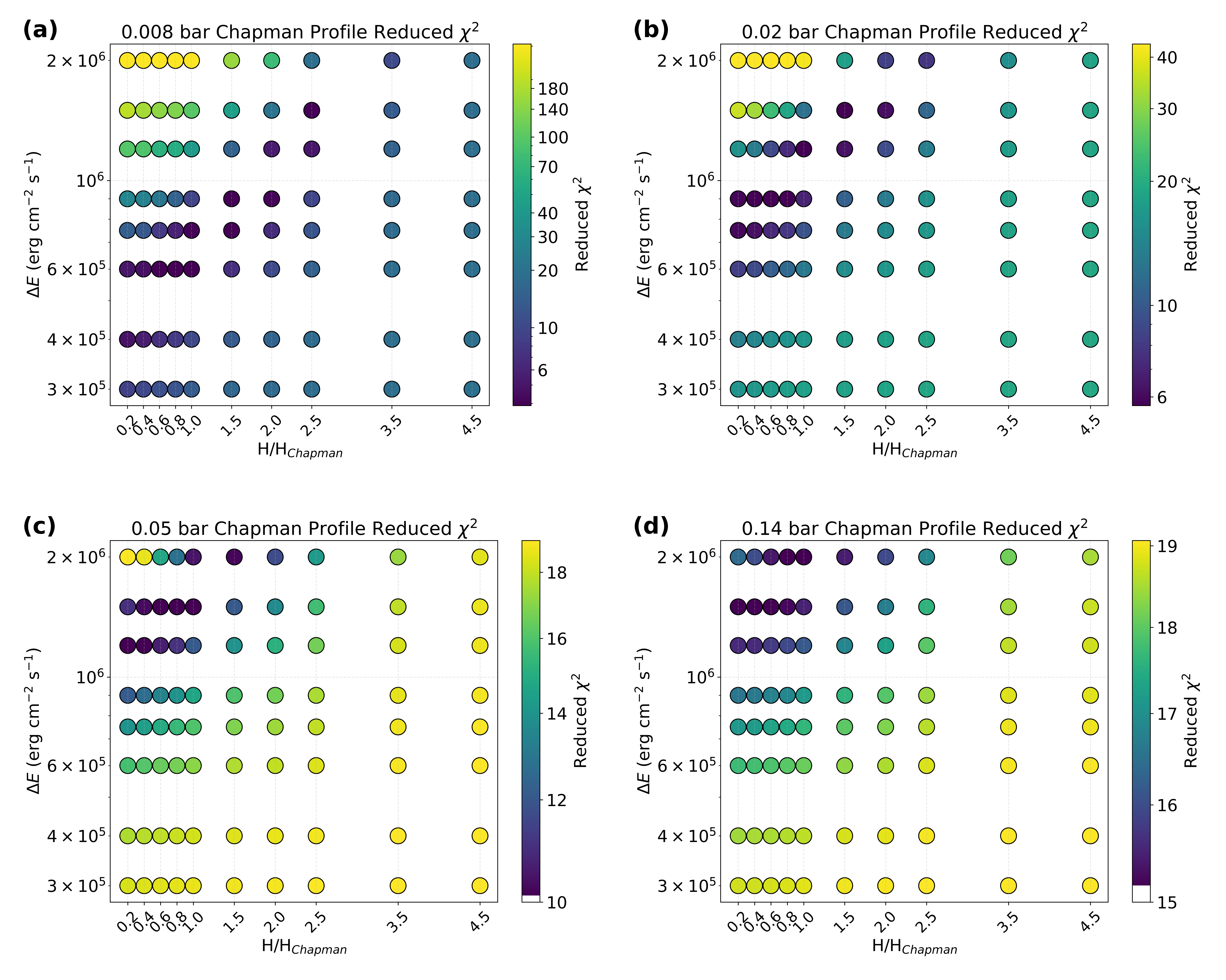}
\caption{Reduced $\chi^2$ values computed from the comparison between the modeled and observed W1935 spectra over the 3.30--3.35~$\mu$m methane emission region for Chapman-profile energy injection models. Each panel corresponds to a different pressure of peak energy deposition: (a) 0.008 bar, (b) 0.02 bar, (c) 0.05 bar, and (d) 0.14 bar. The x-axis shows the Chapman scale-height parameter, $H/H_{\rm Chapman}$, while the y-axis shows the total injected energy, $\Delta E$. Colors indicate the reduced $\chi^2$ of each model, with lower values corresponding to better agreement with the observed methane emission feature. Note that each panel uses an independent color scale (shown by the corresponding color bar) to emphasize the variation in reduced $\chi^2$ within that deposition pressure. The lowest reduced $\chi^2$ values are obtained for energy deposition pressures of 0.008--0.02 bar, whereas models with deeper energy deposition generally provide poorer fits. Larger $H/H_{\rm Chapman}$ values correspond to more vertically confined heating, while smaller values correspond to broader heating profiles.}
\label{fig:1}
\end{figure*}

While several input configurations reproduce portions of the spectrum, the strength and shape of the methane emission feature at $3.325~\mu\mathrm{m}$ are best matched by a limited subset of energy injection profiles that depend on both the pressure level and vertical extent of the heating. In Figure~\ref{fig:1}, we quantify the agreement between the models and the observed spectrum using the reduced $\chi^2$ calculated across the 3.30--3.35~$\mu$m methane emission region. Because our goal is to estimate the total energy required to produce the inversion, we focus on models that reproduce the methane emission feature specifically and not the entire available spectrum of the object. While the best-fit models reproduce the observed methane emission feature, the corresponding reduced $\chi^2$ values remain greater than unity. The JWST spectrum in this wavelength range has very small observational uncertainties, so even modest systematic deviations between the model and data contribute substantially to the total $\chi^2$. Consequently, the reduced $\chi^2$ values should be interpreted primarily as a relative metric for comparing models within the Chapman grid rather than as an indication that the overall fit is poor.

\begin{figure*}[!htb]
\centering
\includegraphics[scale=0.45]{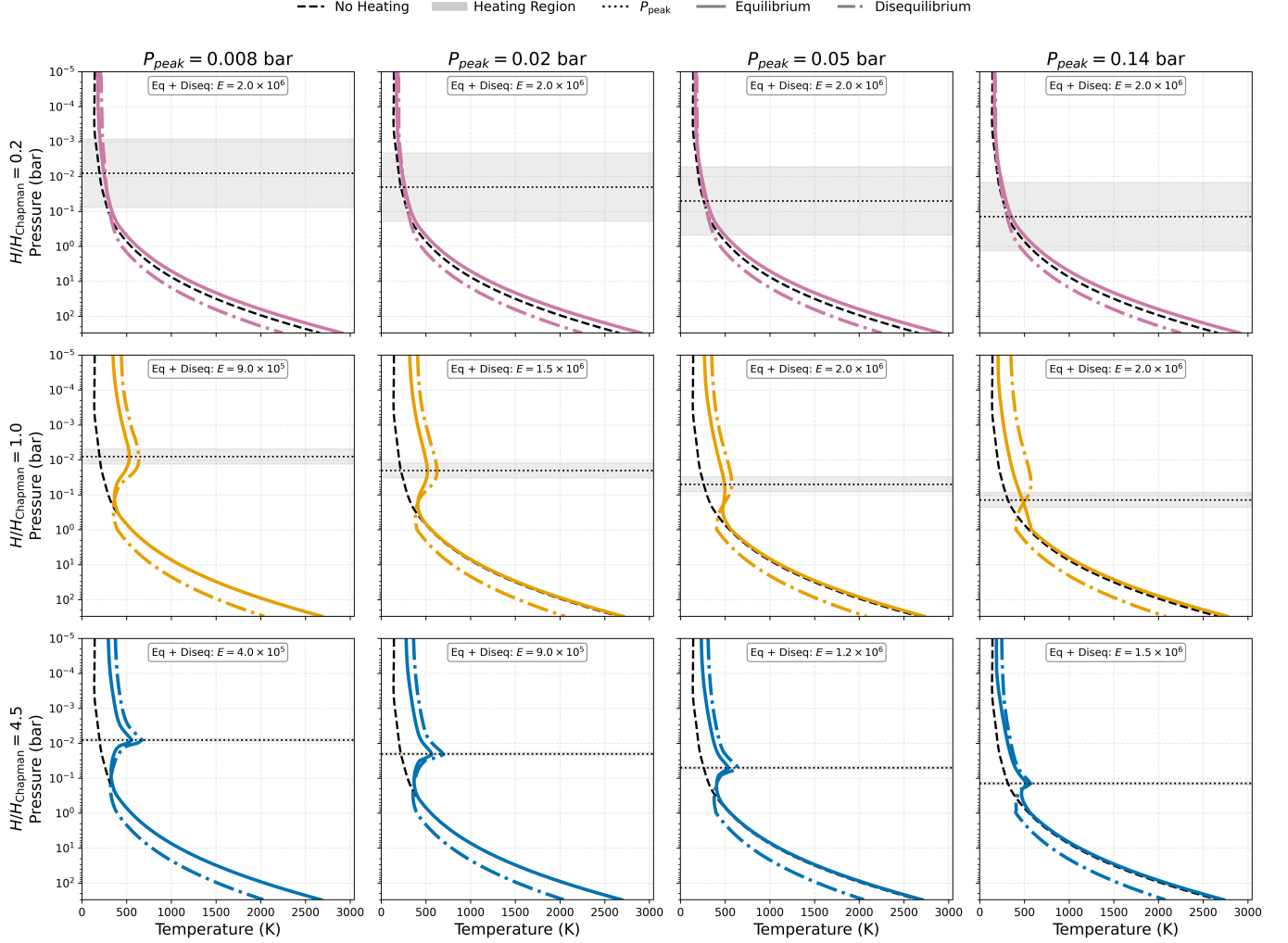}
\caption{Temperature--pressure profiles for representative Chapman heating models spanning the range of peak deposition pressures and heating widths explored in this work. Columns correspond to peak energy deposition pressures of $P_{\rm peak}=0.008$, 0.02, 0.05, and 0.14 bar. Rows correspond to $H/H_{\rm Chapman}=0.2$, 1.0, and 4.5. Thus, the top row represents the most vertically extended (broadest) heating profiles, while the bottom row represents the most vertically concentrated (narrowest) heating profiles. Colored solid curves show the heated atmospheric structures, and the dashed black curve shows the no-heating radiative--convective equilibrium profile. the colored dashed-dotted curves in each panel are the disequilibrium models. The gray shaded regions indicate the characteristic vertical extent of the Chapman heating profile, and the dotted horizontal lines mark the pressure of maximum energy deposition.}
\label{fig:2}
\end{figure*}

\begin{figure*}[!htb]
\centering
\includegraphics[scale=0.45]{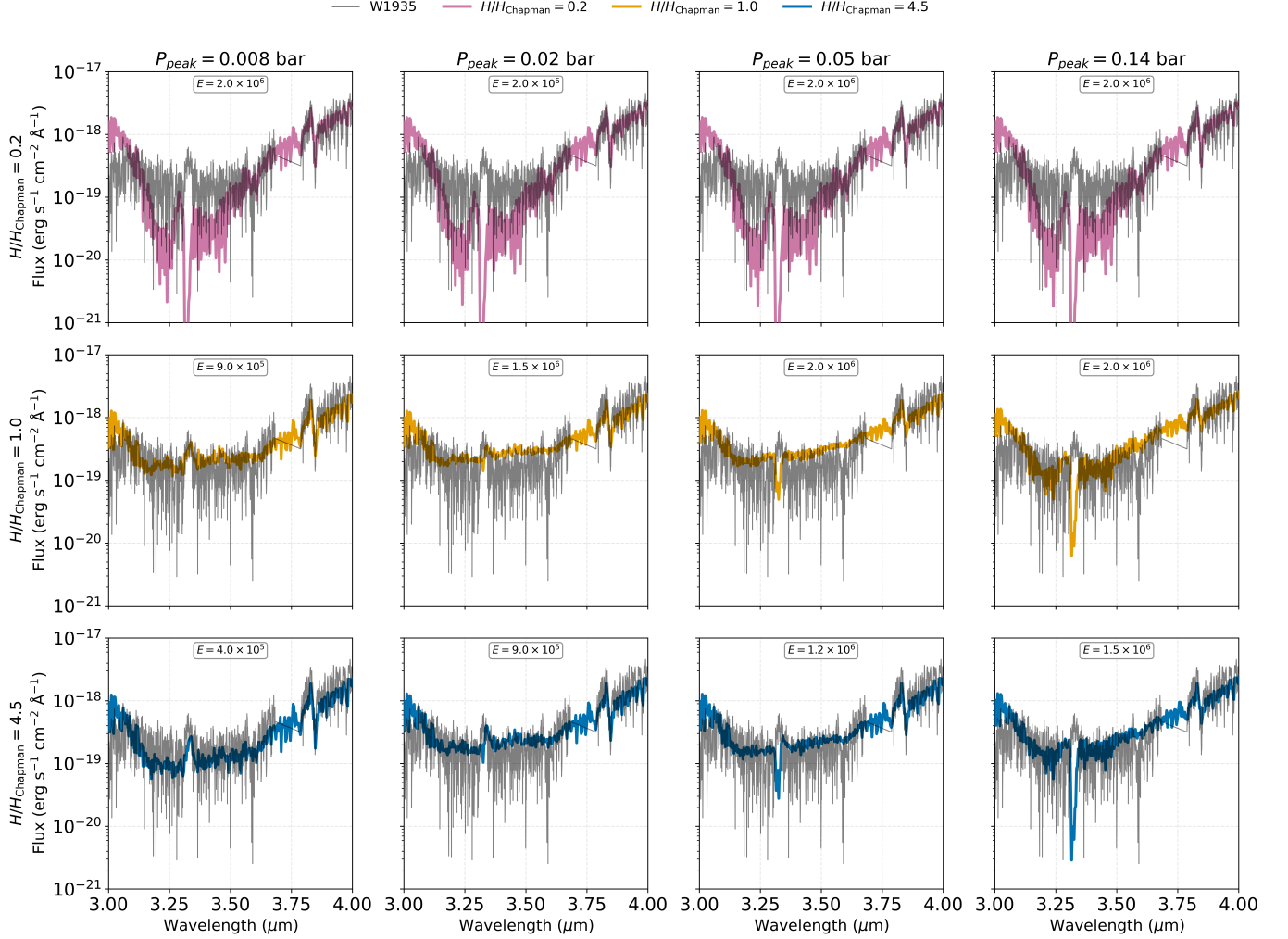}
\caption{Observed spectrum of W1935 (black) compared to representative Chapman-heating models (colored) for four peak deposition pressures ($P_{\rm peak}=0.008$, 0.02, 0.05, and 0.14 bar; columns) and three characteristic heating scale heights ($H/H_{\rm Chapman}=0.2$, 1.0, and 4.5; rows). The plotted models correspond to the minimum reduced $\chi^2$ solutions for each pressure--scale-height combination, where the fit was evaluated over the $3.30$--$3.35~\mu$m methane feature.}
\label{fig:3}
\end{figure*}

Figure~\ref{fig:1} reveals a clear dependence of the model fit on the pressure of peak energy deposition, the total injected energy, and the vertical extent of the heating profile. In general, models with higher-altitude heating ($P_{\rm peak}=0.008$--$0.02$ bar) require less injected energy and provide the best agreement with the observed methane emission feature. At these shallow deposition levels, the best fits occur for intermediate heating widths ($H/H_{\rm Chapman}\sim1.5$--$2.5$), whereas deeper deposition levels ($P_{\rm peak}=0.05$--$0.14$ bar) require both the highest energy injection rates explored and the broadest heating profiles ($H/H_{\rm Chapman}\lesssim1.0$) to minimize the reduced $\chi^2$. Representative temperature--pressure profiles and their corresponding spectra are presented in Figures~\ref{fig:2} and \ref{fig:3}, illustrating how the location, magnitude, and vertical extent of the heating modify both the atmospheric structure and the resulting methane emission.

As the heating is shifted deeper into the atmosphere, the quality of the fit progressively deteriorates and increasingly larger energy injection rates are required to reproduce the observations. This trend suggests that energy deposited closer to the methane-emitting region is more effective at generating the temperature inversion responsible for the emission feature. In contrast, energy deposited at greater depths is redistributed through optically thick atmospheric layers before reaching the upper atmosphere, reducing its ability to directly modify the thermal structure of the methane-forming region. Consequently, deeper heating requires a larger overall energy budget and, at lower injection rates, often fails to produce a sufficiently strong inversion, resulting in spectra dominated by methane absorption rather than emission.

Figure~\ref{fig:2} shows representative temperature--pressure ($T$--$P$) profiles for Chapman heating models spanning the range of energy deposition pressures and heating widths explored in this work. For a fixed deposition pressure, decreasing $H/H_{\rm Chapman}$ increases the vertical extent of the Chapman heating profile, distributing the deposited energy over a broader region of the atmosphere. In contrast, larger values of $H/H_{\rm Chapman}$ concentrate the energy deposition into a narrower range of pressures, resulting in more localized heating. Heating deposited at lower pressures produces the largest temperature enhancements in the methane-emitting region, whereas deeper energy deposition primarily affects higher-pressure layers and requires larger energy fluxes to produce comparable changes in the observable atmosphere. These results demonstrate that both the depth and vertical extent of the deposited energy play important roles in determining the resulting thermal structure.

When energy is deposited at low pressures ($P_{\rm peak}=0.008$--0.02 bar), the heating occurs within the observed methane line forming region, producing a temperature inversion in the observable atmosphere. This inversion drives methane into emission at $3.325~\mu\mathrm{m}$, consistent with the observed spectrum of W1935. In contrast, when the energy is deposited deeper in the atmosphere, the heating remains confined to optically thick layers and has a reduced influence on the observable atmosphere. Consequently, progressively larger energy injection rates are required to modify the upper-atmospheric thermal structure, and the resulting spectra are often dominated by methane absorption rather than emission.

The atmospheric response also depends on the vertical extent of the heating. Although the broadest heating profiles ($H/H_{\rm Chapman}=0.2$) may yield the lowest reduced $\chi^2$ within some regions of parameter space, they generally fail to produce methane emission and instead remain in absorption. Their relatively low $\chi^2$ values therefore reflect an improvement relative to other absorption models rather than a successful reproduction of the observed emission feature. In contrast, the narrower heating profiles ($H/H_{\rm Chapman}=1.0$ and 4.5) more readily produce methane emission, particularly when the heating occurs at lower pressures. This behavior suggests that concentrating the deposited energy within a more localized region of the atmosphere is more effective at producing the thermal structure required to reproduce the observed methane emission feature.

One of the key transitions identified in these models is the shift of the methane feature from emission to absorption. This occurs when the temperature structure reverses such that the methane line-forming region becomes cooler than the layers below, reducing the emergent line intensity relative to the continuum. The location of this transition depends sensitively on both the depth and vertical distribution of the deposited energy. Models with peak heating pressures greater than approximately $0.05,\mathrm{bar}$ generally fail to reproduce the observed methane emission, as the heating primarily affects deeper atmospheric layers while leaving the upper atmosphere comparatively cool. These results demonstrate that the observed methane emission in W1935 is controlled primarily by the amount of energy deposited at low pressures, rather than by deep atmospheric heating alone.

\subsection{Equilibrium versus Disequilibrium}
\label{sec:3.1}
Thus far in this work, W1935 has been modeled under the assumption of chemical equilibrium, in which molecular abundances are determined solely by the local temperature and pressure. In reality, atmospheric transport can drive the composition away from equilibrium by vertically mixing material between atmospheric layers faster than chemical reactions can restore equilibrium abundances. This process, commonly referred to as disequilibrium chemistry, can significantly alter the abundances of key molecules such as CO and CH$_4$ \citep{2020AJ....160...63M}, which strongly influence the thermal structure and emergent spectrum of brown dwarf atmospheres. To assess the impact of disequilibrium chemistry on the atmospheric temperature--pressure structure, emergent spectra, and inferred energy injection requirements of W1935, we perform a disequilibrium chemistry analysis. Specifically, we include vertical mixing through the self-consistent $K{zz}$ method implemented in \texttt{PICASO}, rather than assuming a constant $K_{zz}$ profile, following \cite{2024ApJ...963...73M} and \cite{2026ApJ..1000...98M}.

\begin{figure*}[!htb]
    \centering
    \includegraphics[scale=0.45]{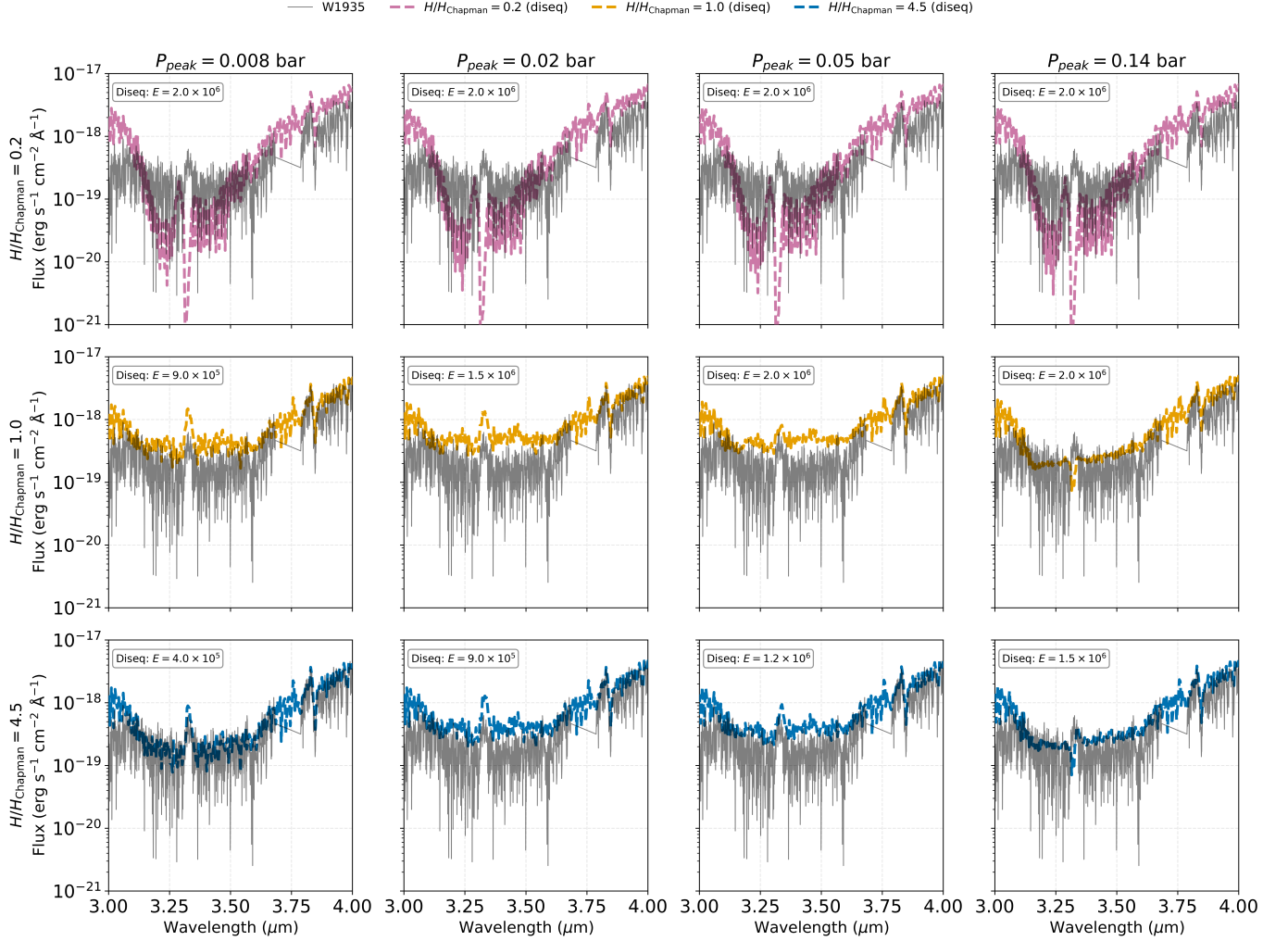}
    \caption{Similar to Figure~\ref{fig:3}, but the colored dashed curves in each panel are the disequilibrium models.}
    \label{fig:5}
\end{figure*}

To assess the impact of disequilibrium chemistry across the full parameter space, we compare the equilibrium and disequilibrium models for the best-fitting cases spanning four pressure levels for peak energy deposition ($P_{\rm peak}=0.008$, 0.02, 0.05, and 0.14 bar) and three representative heating distributions corresponding to $H/H_{\rm Chapman}=0.2$, 1.0, and 4.5. Figure~\ref{fig:2} shows the resulting temperature--pressure profiles, while Figure ~\ref{fig:3} and ~\ref{fig:5} show the corresponding emergent spectra to the observed spectrum of W1935.

Across nearly all cases, the disequilibrium models exhibit warmer temperatures in the upper atmosphere relative to their equilibrium counterparts. The differences are most pronounced for the narrower heating profiles ($H/H_{\rm Chapman}=1.0$ and 5.0), where the disequilibrium temperature structures develop stronger inversions near the energy deposition region, resulting in larger deviations from the equilibrium profiles and enhanced methane emission. In contrast, the broadest heating profile ($H/H_{\rm Chapman}=0.2$) exhibits smaller differences between the equilibrium and disequilibrium temperature structures. Because the deposited energy is distributed over a larger pressure range, the resulting temperature inversions are weak and less sensitive to the effects of chemical quenching, leading to more modest changes in the atmospheric thermal structure.

These differences in thermal structure are reflected in the emergent spectra shown in Figure~\ref{fig:5}. For all four pressure layers, the disequilibrium models generally produce stronger methane emission features than the corresponding equilibrium models for the same atmospheric heating configuration. The enhancement is particularly evident near the 3.325~$\mu$m methane band, where the disequilibrium spectra exhibit increased emission relative to equilibrium models. This behavior suggests that vertical mixing alters the atmospheric opacity structure--and consequently the efficiency of radiating away the deposited energy--in a manner that allows added energy to more effectively influence the pressures probed by the methane feature in the visible region of the atmosphere. 

The magnitude of the disequilibrium effect depends on both the pressure of peak energy deposition and the vertical extent of the heating profile. At shallow deposition pressures ($P_{\rm peak}=0.008$--0.02 bar), the equilibrium and disequilibrium spectra are often similar because the heating is already deposited close to the methane-emitting region. However, as the heating is moved deeper into the atmosphere ($P_{\rm peak}=0.05$--0.14 bar), disequilibrium chemistry becomes increasingly important. In these cases, the disequilibrium models are able to reproduce comparable methane emission strengths with substantially less injected energy than required in the equilibrium calculations.

\begin{figure*}[!htb]
    \centering
    \includegraphics[scale=0.45]{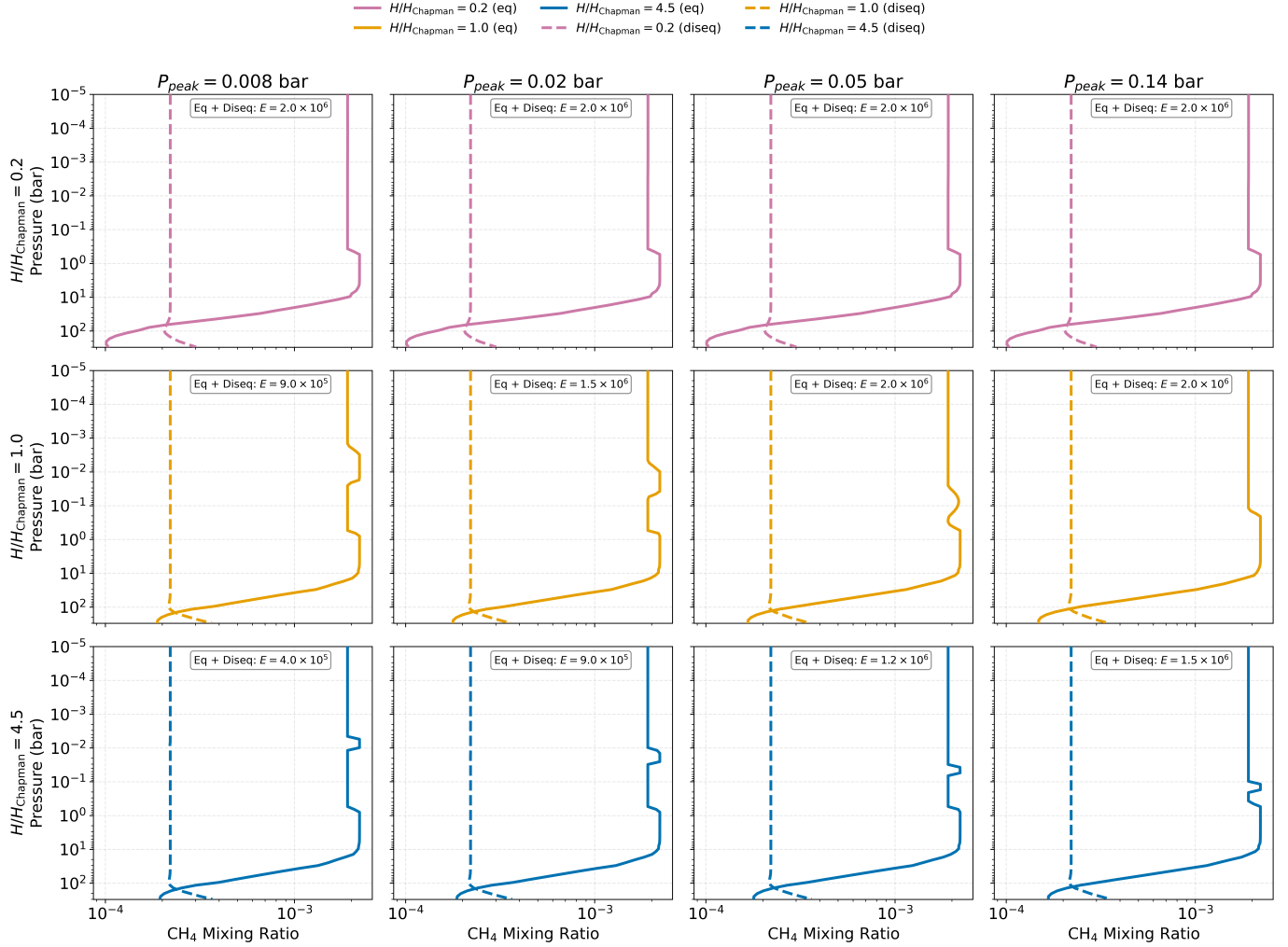}
    \caption{Mixing ratio grid as a function of pressure of methane between the equilibrium and disequilibrium models. Rows and columns are the same as in Figures~\ref{fig:2}, \ref{fig:3}, and \ref{fig:5}. The equilibrium CH$_4$ abundance is significantly altered by the presence of a temperature inversion, while the disequilibrium abundance remains largely quenched throughout the atmosphere.}
    \label{fig:ch4}
\end{figure*}

Taken together, these results demonstrate that the inferred heating requirements depend sensitively on the chemical state of the atmosphere. Incorporating disequilibrium chemistry consistently increases the efficiency with which deposited energy generates observable methane emission, reducing the total energy input required to reproduce the spectrum of W1935. While the exact reduction varies across the parameter space, the overall trend is robust: models that include vertical mixing and chemical quenching generally require less heating than equilibrium models to produce the observed emission feature. This is likely because the lower methane abundance under disequilibrium conditions radiatively cools the atmosphere less efficiently, requiring less energy input to produce the same temperature inversion. As shown in Figure~\ref{fig:ch4}, the equilibrium CH$_4$ mixing ratio is more sensitive to the presence of a temperature inversion, whereas the disequilibrium CH$_4$ mixing ratio remains largely quenched throughout most of the atmosphere. A more detailed exploration of this effect is saved for future work.


\subsection{Prediction of Spectra from $5$--$15\,\mu$m}
\label{sec:3.2}
The spectrum presented in \citep{Faherty_2024} covers only $\sim$3--5~$\mu$m, with longer-wavelength \textit{JWST} observations not yet publicly available (JWST GO 7793). In this section, we use \texttt{PICASO} to predict the longer-wavelength spectral appearance of W1935 from $5$--$15\,\mu$m for the two different best-fit atmospheric energy-deposition pressure cases in this study. A prediction over this wavelength range is also presented in \citep{suarez2025diversitycoldworldspredicted}. 

\begin{figure*}[!htb]  
    \centering
    \includegraphics[scale=0.45]{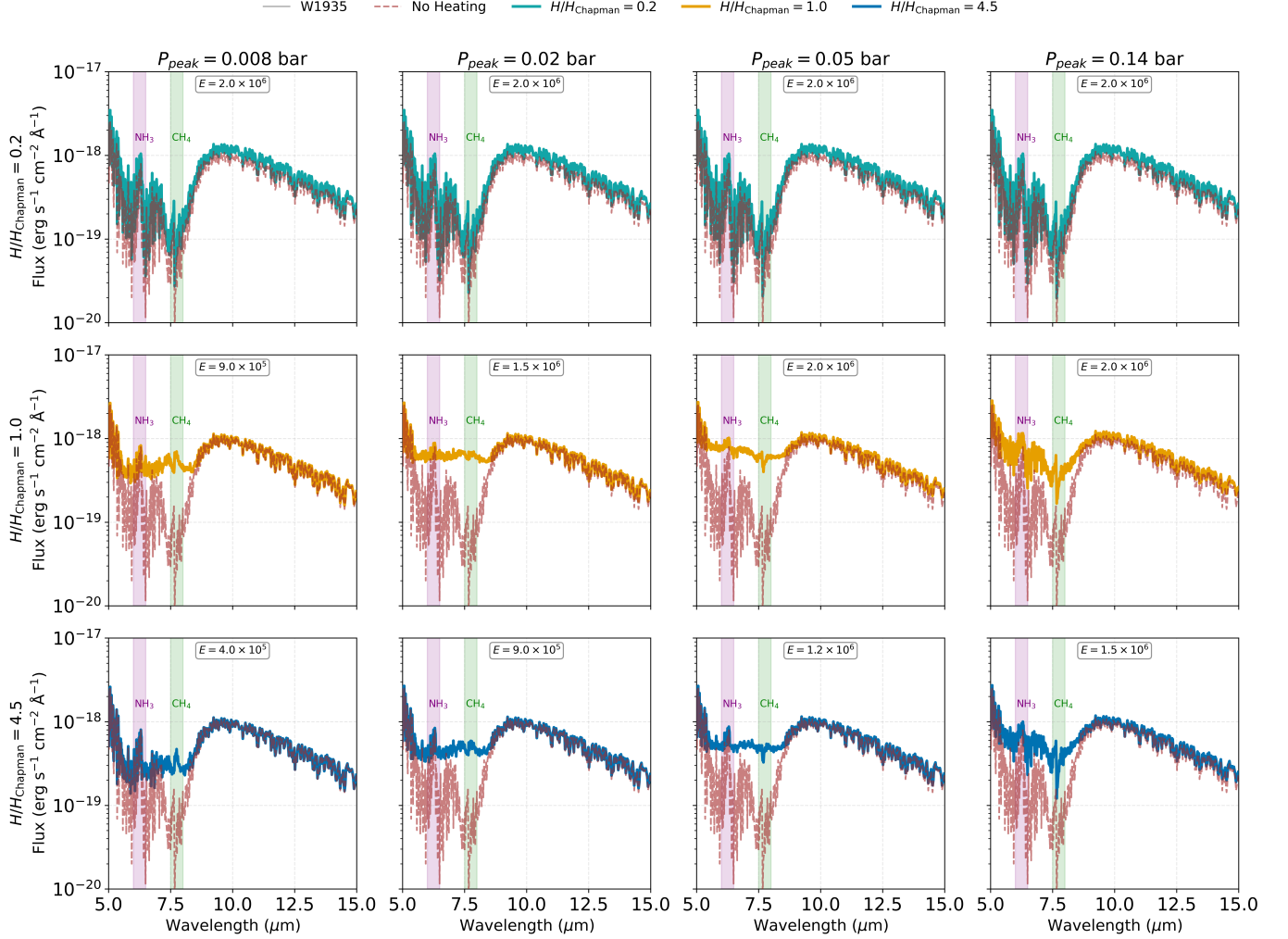}
    \caption{Spectra from 5--15~$\mu$m for the best-fitting heating models across the explored parameter space. Columns correspond to the pressure of peak energy deposition ($P_{\rm peak}=0.008$, 0.02, 0.05, and 0.14~bar), while rows correspond to the vertical extent of the heating profile ($H/H_{\rm Chapman}=0.2$, 1.0, and 4.5). The models shown represent the best-fitting energy injection rates determined from the reduced $\chi^2$ analysis in Figure~\ref{fig:1}. Variations in both the deposition pressure and heating distribution modify the strength of molecular absorption and emission features, illustrating how the thermal structure responds to different heating geometries. Molecular features are labeled following \cite{2006ApJ...648..614C}.} 
    \label{fig:6}
\end{figure*}

Figure~\ref{fig:6} presents the 5--15~$\mu$m spectra of the best-fitting equilibrium models identified from the 3.325~$\mu$m methane analysis. The columns correspond to different pressures of peak energy deposition ($P_{\rm peak}=0.008$, 0.02, 0.05, and 0.14~bar), while the rows show different vertical extents of the heating profile ($H/H_{\rm Chapman}=0.2$, 1.0, and 4.5). With the exception of the parameter-space configurations that fail to produce a temperature inversion or generate only a weak inversion (Figure~\ref{fig:2}), the inclusion of energy deposition produces a thermal inversion that drives methane emission near $7.8~\mu$m. In contrast, the corresponding no-heating models exhibit methane in absorption at these wavelengths. The persistence of this emission feature across nearly the entire model grid indicates that the thermal inversions required to reproduce the observed 3.325~$\mu$m methane emission also generate observable signatures in the mid-infrared, providing an additional diagnostic of upper-atmospheric heating.

The spectra demonstrate that both the depth and vertical extent of the heating influence the strength and morphology of the resulting molecular features. For the broadest heating profile ($H/H_{\rm Chapman}=0.2$), the NH$_3$ and CH$_4$ bands generally remain in absorption across all deposition pressures, indicating that distributing the deposited energy over a large pressure range is insufficient to produce a strong temperature inversion in the observable atmosphere. In contrast, the narrower heating profiles ($H/H_{\rm Chapman}=1.0$ and 4.5) more readily produce molecular emission, particularly when the heating is deposited at lower pressures. As the deposition pressure increases to $P_{\rm peak}=0.05$ and $0.14$~bar, the emission features weaken, consistent with a larger fraction of the injected energy being deposited below the photosphere. These results demonstrate that both the depth and vertical extent of the heating play important roles in determining whether molecular bands appear in emission or absorption. 

A complementary response is observed in the NH$_3$ feature near $6$--$6.5~\mu$m. As the heating profile becomes more effective at producing a thermal inversion, the ammonia absorption feature weakens relative to the no-heating case and, in some models, transitions toward emission. This behavior reflects changes in the temperature structure of the upper atmosphere and the strength of the inversion. Together, the NH$_3$ and CH$_4$ features provide complementary diagnostics of atmospheric heating, with methane tracing the development of the inversion and ammonia probing the resulting thermal structure. These results suggest that future JWST/MIRI observations could help distinguish between different heating scenarios by simultaneously constraining the strengths and morphologies of the ammonia and methane features in the mid-infrared.

\subsection{Fractional Area Analysis}
\label{sec:3.3}
In Saturn's atmosphere, the strongest stratospheric thermal emission is observed by ground-based observations near the poles during the summer \citep[]{Fletcher_Greathouse_Guerlet_Moses_West_2018}. If the methane emission from W1935 is similarly localized then more energy would be required over an area smaller than the full disk of the brown dwarf. To evaluate how a reduced emitting area impacts the required energy deposition in W1935’s atmosphere, we performed a fractional flux analysis. This approach tests how emission from a smaller effective surface area modifies the inferred energy requirements. The results are presented in Table~\ref{tab:3}. To do this, the following equation was used:

\begin{equation}
F_{\lambda,\mathrm{mix}}
=
f_{\mathrm{emit}}\,F_{\lambda,\mathrm{hot}}
+
\left(1-f_{\mathrm{emit}}\right)\,F_{\lambda,\mathrm{base}}
\end{equation}

\noindent where \(F_{\lambda}\) is the emergent flux of the combined atmosphere, \(F_{\lambda,\mathrm{hot}}\) is the flux from the heated region, \(F_{\lambda,\mathrm{base}}\) is the flux from the unheated atmosphere, and \(f_{\mathrm{emit}}\) is the fractional area of the brown dwarf surface covered by the heated region. For example, if \(f_{\mathrm{emit}} = 0.6\), then 60\% of the visible surface contributes flux from the heated atmosphere, while the remaining 40\% contributes flux from the baseline atmosphere. As \(f_{\mathrm{emit}}\) decreases, a smaller fraction of the observed disk contributes the enhanced emission, requiring a larger energy injection to reproduce the observed methane emission feature.

\begin{table*}[htbp]
\centering
\caption{Best-fitting injected energy requirements for different emitting fractions ($f$) and heating profile widths at $P_{\rm peak}=0.008$~bar. The ``fractional'' column gives the energy required when only a fraction $f$ of the observed disk contributes to the heated spectrum, while the ``100\%'' column gives the corresponding flux fraction for an atmosphere emitting across 100\% of the observed disk. Note that 5.00~$\times$~10$^6$ erg~cm$^{-2}$~s$^{-1}$ is the upper bound of the explored energy grid.}
\label{tab:3}

\begin{tabular}{ccccc}
\hline
$P_{\rm peak}$ (bar) &
$H/H_{\rm Chapman}$ &
$f$ &
Injected Energy (Fractional) &
Injected Energy (100\%) \\
&
&
&
($\mathrm{erg\,cm^{-2}\,s^{-1}}$) &
($\mathrm{erg\,cm^{-2}\,s^{-1}}$) \\
\hline

0.008 & 0.2 & 0.1 & $5.00\times10^{6}$ & $2.00\times10^{6}$ \\
0.008 & 1.0 & 0.1 & $3.97\times10^{6}$ & $9.00\times10^{5}$ \\
0.008 & 4.5 & 0.1 & $3.16\times10^{6}$ & $4.00\times10^{5}$ \\
\hline

0.008 & 0.2 & 0.2 & $5.00\times10^{6}$ & $2.00\times10^{6}$ \\
0.008 & 1.0 & 0.2 & $2.51\times10^{6}$ & $9.00\times10^{5}$ \\
0.008 & 4.5 & 0.2 & $1.99\times10^{6}$ & $4.00\times10^{5}$ \\
\hline

0.008 & 0.2 & 0.4 & $5.00\times10^{6}$ & $2.00\times10^{6}$ \\
0.008 & 1.0 & 0.4 & $1.99\times10^{6}$ & $9.00\times10^{5}$ \\
0.008 & 4.5 & 0.4 & $1.58\times10^{6}$ & $4.00\times10^{5}$ \\
\hline

0.008 & 0.2 & 0.8 & $5.00\times10^{6}$ & $2.00\times10^{6}$ \\
0.008 & 1.0 & 0.8 & $1.26\times10^{6}$ & $9.00\times10^{5}$ \\
0.008 & 4.5 & 0.8 & $7.11\times10^{5}$ & $4.00\times10^{5}$ \\
\hline

\end{tabular}
\end{table*}

To investigate the effect of partial surface coverage, we computed a grid of models with energy deposition centered at $0.008$~bar and heating profile widths corresponding to $H/H_{\rm Chapman}=0.2$, 1.0, and 4.5. For each configuration, the injected energy was varied over logarithmically spaced steps and the best-fitting value was determined by minimizing the reduced $\chi^2$ of the methane emission feature. The resulting energy requirements are summarized in Table~\ref{tab:3}.

In all cases, the required energy decreases as the emitting fraction increases. This behavior is expected because a larger fraction of the observed disk contributes to the heated spectrum, reducing the local heating rate needed to reproduce the observed methane emission feature. For the intermediate-width profile ($H/H_{\rm Chapman}=1.0$), the best-fitting energy decreases from $3.97\times10^{6}$ to $1.26\times10^{6}\ \mathrm{erg\ cm^{-2}\ s^{-1}}$ as the emitting fraction increases from $f=0.1$ to $f=0.8$. A similar trend is seen for the narrowest heating profile ($H/H_{\rm Chapman}=4.5$), where the required energy decreases from $3.16\times10^{6}$ to $3.42\times10^{5}\ \mathrm{erg\ cm^{-2}\ s^{-1}}$.

The dependence on heating profile width is also significant. At fixed emitting fraction, the narrower heating profiles generally require less energy to reproduce the observed methane emission feature because the deposited energy is concentrated within a smaller pressure range, producing a stronger local temperature inversion. In contrast, the broadest heating profile ($H/H_{\rm Chapman}=0.2$) consistently requires the largest energy input, reaching the upper bound of the explored energy grid for all emitting fractions. Despite this substantial energy deposition, these models fail to produce methane emission and instead remain dominated by absorption features. These results suggest that concentrating the deposited energy within a relatively narrow pressure range in the atmosphere is more effective at generating the thermal structure required to reproduce the observed spectrum of W1935.

For comparison, the fully emitting ($f=1$) models require injected energies of $2.0\times10^{6}$, $9.0\times10^{5}$, and $4.0\times10^{5}\ \mathrm{erg\ cm^{-2}\ s^{-1}}$ for $H/H_{\rm Chapman}=0.2$, 1.0, and 4.5, respectively. These results demonstrate a degeneracy between the emitting fraction and the local heating rate: a smaller emitting area can reproduce the observed methane emission feature, but only at the expense of substantially larger energy deposition within the heated region. The results further indicate that narrower heating profiles are more efficient at producing methane emission, requiring lower energy fluxes than broader heating profiles for a given emitting fraction.

\section{Potential Heating Mechanisms}
\label{sec:4}
The heating mechanism driving the temperature inversion and methane emission on W1935 is unknown at present. \citet{Faherty_2024} suggested that heating by auroral processes is a plausible explanation for the temperature inversion but did not present specific calculations to support this hypothesis. Instead, their argument was based on the observation of radio emissions attributed to auroral processes from some brown dwarfs \citep{hallinan06,kao18} and a qualitative comparison with the upper atmosphere of Jupiter where auroral heating is the likely explanation for the relatively high temperatures in the thermosphere ($p \lesssim$~10$^{-6}$~bar) \citep{mullerwodarg25}. The so-called energy crisis on the giant planets i.e., the fact that the temperatures are much higher than expected based on solar heating only, is limited to the thermosphere on Jupiter, Saturn, and Neptune. Only on Uranus does it extend to the middle atmosphere \citep[e.g.,][]{marley99}, similar in pressure range to the temperature inversion on W1935. In that case, however, the likely solution in the middle atmosphere is absorption of sunlight by stratospheric aerosols \citep{milcareck24} while auroral processes might help to heat the thermosphere \citep{gershman25}. 


The proposed solutions to the energy crisis in giant planet upper atmospheres include upward propagating gravity or sound waves \citep{young97,schubert03}, wind-driven (dynamo) electrodynamics or plasma waves \citep[e.g.,][]{smith13}, and auroral heating \citep[e.g.,][]{brown20,mullerwodarg25}. While gravity wave breaking could be a substantial energy source on W1935, \citet{Faherty_2024} argues that gravity waves are ubiquitous and if their heating impact was significant, similar temperature inversions would probably be more common in brown dwarf atmospheres. W1935 may have the right atmospheric/physical parameters where wave-breaking could be depositing energy at precisely the pressure range needed to sustain the observed inversion, a phenomenon that may not be common among similar objects. In the Solar System context, heating by atmospheric waves is now also thought to be insufficient to explain the observed temperatures in the thermospheres of Jupiter and Saturn \citep{strobel18,mullerwodarg19}. Furthermore, quantifying the net thermal impact of gravity wave breaking requires self-consistent dynamical modeling because wave dissipation deposits both momentum and energy, making the resulting atmospheric heating difficult to estimate with simple order-of-magnitude calculations \citep{2022GeoRL..4997219B}. A detailed assessment of gravity-wave heating is therefore beyond the scope of the present work.

Similar considerations apply to dynamo electric fields and plasma waves that are further limited by the very low inherent plasma densities in the middle atmosphere of W1935 (see Section~\ref{sec:4.1}). This leaves auroral heating as the option to explore further from the solar system context. Below, we also consider the influx of cometary material or other external material as a possible energy source. 

\subsection{Prospects for Joule Heating}
\label{sec:4.1}
As we note above, auroral heating is a plausible explanation for the temperatures observed in the thermospheres of Jupiter and Saturn. In both cases, however, the direct energy input from precipitating auroral electrons is negligible and auroral heating is actually dominated by resistive Joule heating \citep[e.g.,][]{mullerwodarg19,mullerwodarg25}. Joule heating refers to the frictional heating of the neutral atmosphere by collisions with perpendicular electric currents that close magnetic field-aligned auroral currents in the ionosphere. The field-aligned currents connect the ionosphere to regions of the magnetosphere where processes associated with either solar wind interaction or internal plasma populations are driving the aurora. For Joule heating (by auroral currents or dynamo currents) or other plasma effects to be significant, the plasma frequency must exceed the electron-neutral collision frequency. Typically, it is also assumed that the ion gyrofrequency must exceed the ion-neutral collision frequency because fundamentally Joule heating arises from collisional friction between the plasma and the neutral atmosphere, indicating that ions should be decoupled from the neutral atmosphere.

\begin{figure*}[!htb]  
    \centering
    \includegraphics[scale=1.1]{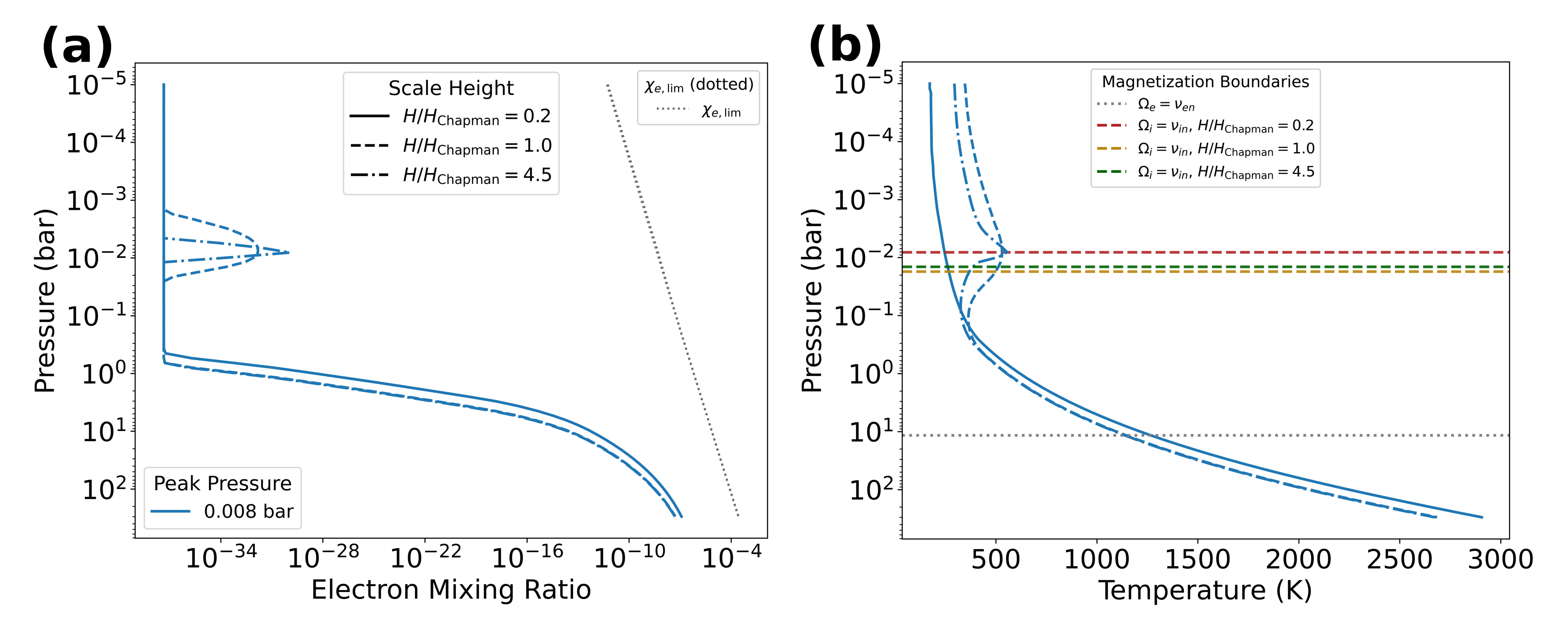}
    \caption{Model electron mixing ratio profiles and temperature--pressure structures for the best-fitting Chapman heating models of W1935 with $P_{\rm peak}=0.008$~bar. Line styles correspond to $H/H_{\rm Chapman}=0.2$, 1.0, and 4.5. Panel~(a) shows the electron mixing ratio profiles, while the gray dotted curve indicates the critical electron mixing ratio required for electron magnetization, $\chi_{e,\mathrm{lim}}$ (corresponding to $\Omega_e=\nu_{en}$). Panel~(b) shows the corresponding temperature--pressure profiles, with colored horizontal dashed lines indicating the ion magnetization boundaries ($\Omega_i=\nu_{in}$) for each heating profile. Magnetization boundaries were calculated assuming a dipolar magnetic field strength of $8.67$~T ($86.7$~kG).}
    \label{fig:7}
\end{figure*}

The first condition is expressed as:
\begin{equation}
\Omega_p = \sqrt{\frac{n_e e^2}{\epsilon_0 m_e}} = \nu_{en}
\end{equation}
where $\Omega_p$ is the plasma frequency, $n_e$ is the electron density, $e$ is the electron charge, $\epsilon_0$ is the permittivity of free space, $m_e$ is the electron mass, and $\nu_{en}$ is the electron-neutral collision frequency. This leads to lower limit on the electron volume mixing ratio $x_e$ that is a necessary (but not sufficient) condition for plasma effects to be significant:
\begin{equation}
x_e = \frac{\epsilon_0 m_e}{ne^2} \nu_{en}^2
\end{equation}
where $n$ is the total density of the atmosphere. Here, we calculated the electron-H, electron-H$_2$, and electron-He collision frequencies using equations (12) in \citet[]{2010ApJ...722..178K} and density output from our model of W1935. These collision frequencies are important for determining the degree of atmospheric magnetization and the efficiency of plasma processes in brown dwarf atmospheres. Figure~\ref{fig:7} compares the required $x_e$ (dashed line) with the electron mixing ratio based on thermal ionization by our models for W1935 (blue solid, dash, and dash-dot lines). The electron mixing ratio in the atmosphere of W1935 is orders of magnitude too small for plasma effects to play any role in controlling the middle and upper atmosphere unless there is a substantial external ionization source from, say, auroral particle precipitation \citep{2026ApJ...999...26Z}.

The second condition requires that ion mobility exceeds unity:
\begin{equation}
k_{in} = \frac{eB}{m_i \nu_{in}} > 1
\end{equation}
where $m_i$ is the ion mass and $\nu_{in}$ is the ion-neutral collision frequency. We calculate the magnetic field strength $B$ for a centered dipole from
\begin{equation}
B = \frac{M_B}{R_p^3} \sqrt{1 + 3\cos^2\theta_m}
\end{equation}
where $M_B$ is the magnetic dipole moment, $R_p = 0.92\,R_J$ is the radius of W1935, and $\theta_m = 45^\circ$ is an arbitrary magnetic latitude. We use Na$^+$ as the test ion and calculate the ion-neutral collision frequencies with H$_2$, He, and H using \citep[]{SchunkNagy2000}:
\begin{equation}
\nu_{in} = 2.21\pi \frac{n_n}{m_n} \frac{m_i + m_n}{\sqrt{\gamma_n}} \frac{e^2}{m_{in}} \quad \text{(all units in cgs)}
\end{equation}
where $n_n$ is the number density of the neutral collision partner, $m_n$ is the neutral mass, $m_i$ is the ion mass, $\gamma_n$ is the neutral polarizability and $m_{in}$ is the reduced mass of the ion and the neutral species. The bottom of the dynamo layer, on the other hand, is the location where electron mobility exceeds unity: 
\begin{equation}
k_{en} = \frac{eB}{m_e \nu_{en}} > 1.
\end{equation}
The bottom of the dynamo layer is always below the $k_{in} =$~1 level because $k_{en} > k_{in}$. Provided that Joule heating is sufficiently significant in the dynamo layer, this provides a less stringent condition for Joule heating to be important.

Using the Jovian magnetic dipole moment of $M_{BJ} =$~1.56~$\times$~10$^{20}$~T~m$^3$ for W1935 (surface $B =$~867 $\mu$T at $\theta_m =$~45$^{\circ}$) puts the bottom of the dynamo layer at around $p =$~1 mbar and the $k_{in} =$~1 level at lower pressures. This would not allow for significant Joule heating in the middle atmosphere at pressures of 0.0001--0.1 bar where the temperature inversion is observed. Multiplying the magnetic moment by a factor of 100 moves the bottom of the dynamo layer to 0.1 bar (surface $B =$~0.0867 $\mu$T at $\theta_m =$~45$^{\circ}$) and multiplying it by 10,000 moves the $k_{in} =$~1 level to 0.01--0.1 bar (surface $B =$~8.67 T at $\theta_m =$~45$^{\circ}$). Thus dipole magnetic moments of 100--10,000 $M_{BJ}$ may allow for significant Joule heating in the middle atmosphere. It should be noted, however, that much higher electron densities than predicted from thermal ionization are still required to make this possible. Those electron densities, along with Joule heating, would be limited to auroral regions. Whether energy can be redistributed globally to allow for higher CH$_4$ emissions in spatially unresolved observations is an additional open question, even if auroral heating provides enough energy in principle to explain the temperature inversion in the middle atmosphere. 

Recent detections of auroral radio emission from late-L and T dwarfs \citep{2018ApJS..237...25K} indicate kilogauss-strength magnetic fields that exceed predictions from standard dynamo scaling laws and show little dependence on age. These results suggest that unusually strong magnetic fields may be common at the substellar–planetary boundary. In this context, the large magnetic moments (100--10,000 $M_{BJ}$) required in our models are not purely speculative, but are consistent with emerging observational constraints. 

\subsection{Cometary Impact}
\label{sec:4.2}

Although there is currently no observational evidence that W1935 hosts a substantial reservoir of cometary material, cometary impacts provide a useful order-of-magnitude comparison for assessing whether impacts could supply the inferred atmospheric energy budget. Bombardment is well known throughout the Solar System and can deliver substantial energy to both terrestrial and giant planets. For example, \citet{Takata_1995} estimated that the Shoemaker--Levy~9 fragments, with diameters ranging from 0.5--1.8 km, released a total impact energy of $1.2 \times 10^{30}$ erg upon colliding with Jupiter.

To estimate the types of comets capable of delivering the required energy flux of to W1935, we compute the total power required over the surface of the brown dwarf, $4\pi R_{\rm BD}^{2} F$. Assuming impacts occur at the escape velocity, the kinetic energy of a comet of radius and density is $E_{\rm comet} = \frac{1}{2}\left(\frac{4}{3}\pi r_{\rm comet}^{3}\rho_{\rm comet}\right)v_{\rm esc}^{2}$. The required impact rate is therefore

\begin{equation}
\Gamma = \frac{4\pi R_{\rm BD}^{2} F}{E_{\rm comet}} \, ,
\end{equation}

\noindent with the corresponding time between impacts given by $t_{\rm impact} = 1/\Gamma$. By varying the comet radius \(r_{\rm comet}\) and density \(\rho_{\rm comet}\), we can explore how these parameters affect the required impact rate and the time between impacts. A range of radii from 1~km to 20~km is considered, and densities of 0.3, 0.5, and $1.0~\mathrm{gcm^{-3}}$. Considering that the mass of W1935 is not well constrained, we did this analysis for 6, 20, and 35 $M_{\text{Jup}}$.

\begin{figure*}[!htb]  
    \centering
    \includegraphics[scale=0.45]{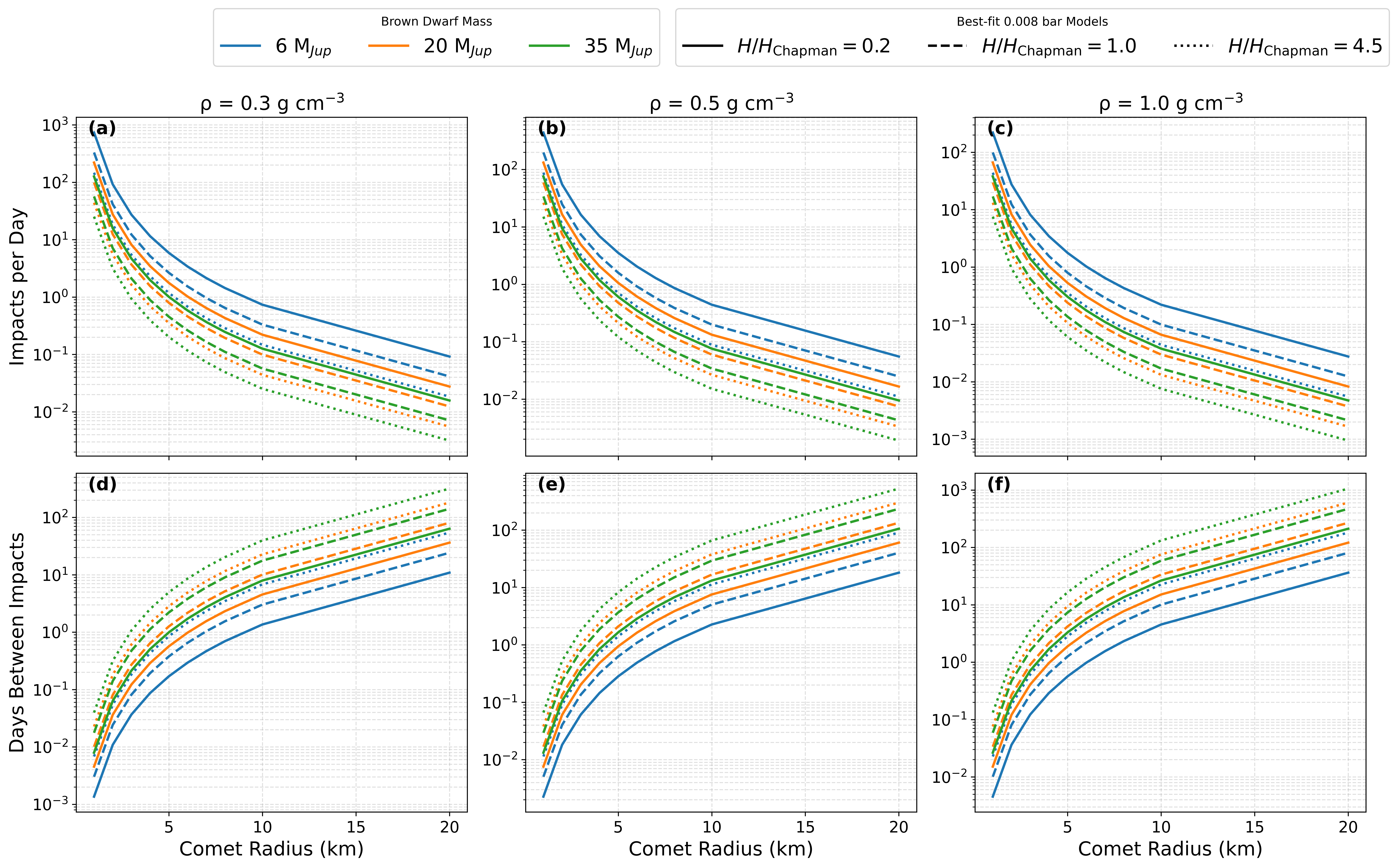}
    \caption{Comet impact rates required to reproduce the best-fitting $P_{\rm peak}=0.008$ bar heating models. The top row shows impacts per day and the bottom row shows days between impacts as a function of comet radius. Columns correspond to different comet densities, colors indicate brown dwarf mass, and line styles denote the three best-fitting heating profiles.}
    \label{fig:8}
\end{figure*}

As noted above, this calculation represents an order-of-magnitude estimate that considers only the total impact energy available. It does not account for the altitude at which that energy is deposited, nor whether impacts preferentially deliver their energy to the $\sim$1--10 mbar region required by our atmospheric models. Figure~\ref{fig:8} therefore illustrates the impact rates necessary to match the inferred energy budget, rather than demonstrating that cometary impacts can reproduce the required atmospheric heating profile.

For representative impact parameters, a 5 km comet with a bulk density of $0.3~\mathrm{g,cm^{-3}}$ impacting a $6~M_{\rm Jup}$ brown dwarf delivers $\sim1.7\times10^{31}~\mathrm{erg}$ per collision. To sustain the heating inferred for the best-fitting $P_{\rm peak}=0.008$ bar models, the required impact frequency ranges from approximately one impact every 0.17 days for the $H/H_{\rm Chapman}=0.2$ case, to every 0.38 days for $H/H_{\rm Chapman}=1.0$, and every 0.85 days for $H/H_{\rm Chapman}=4.5$. Increasing the comet density increases the kinetic energy delivered per impact and correspondingly reduces the required impact frequency. For example, increasing the density from $0.3$ to $1.0~\mathrm{g,cm^{-3}}$ increases the impact energy by a factor of $\sim3$, reducing the required frequency for the $H/H_{\rm Chapman}=0.2$ case from one impact every $\sim0.17$ days to one every $\sim0.57$ days.

The required impact frequency also depends strongly on brown dwarf mass. More massive brown dwarfs have higher escape velocities, resulting in substantially more energetic impacts. For the same 5 km, $\rho=0.3~\mathrm{g,cm^{-3}}$ comet, the time between impacts increases from $\sim0.17$ days for a $6~M_{\rm Jup}$ brown dwarf to $\sim0.57$ days for $20~M_{\rm Jup}$ and nearly one day for $35~M_{\rm Jup}$ in the highest-energy heating scenario. Across the full range of masses, densities, and heating profiles explored here, the required impact rates span from multiple impacts per day to approximately one impact every few years for the largest and densest comets.

Overall, increasing the comet size, density, or brown dwarf mass reduces the required impact frequency by increasing the kinetic energy delivered per collision. However, matching the required energy budget alone is insufficient to explain the observed thermal inversion. Our models indicate that heating must be deposited primarily within the $\sim$1--10 mbar region, and it remains unclear whether cometary impacts naturally deliver most of their energy at these pressures. Moreover, sustaining the inferred heating would require both a substantial reservoir of cometary material and an efficient mechanism for continuously delivering impactors into the atmosphere, raising the question of why W1935 would experience such frequent impacts while most brown dwarfs do not exhibit similar methane emission. Our cometary analysis also neglects atmospheric entry physics and the effects of impact-delivered material on atmospheric chemistry, opacity, and radiative transfer. A complete assessment of the cometary hypothesis will therefore require coupling impact energetics with those atmospheric processes.

\section{Discussion}
\label{sec:5}

\begin{figure*}[!htb]  
    \centering
    \includegraphics[scale=0.95]{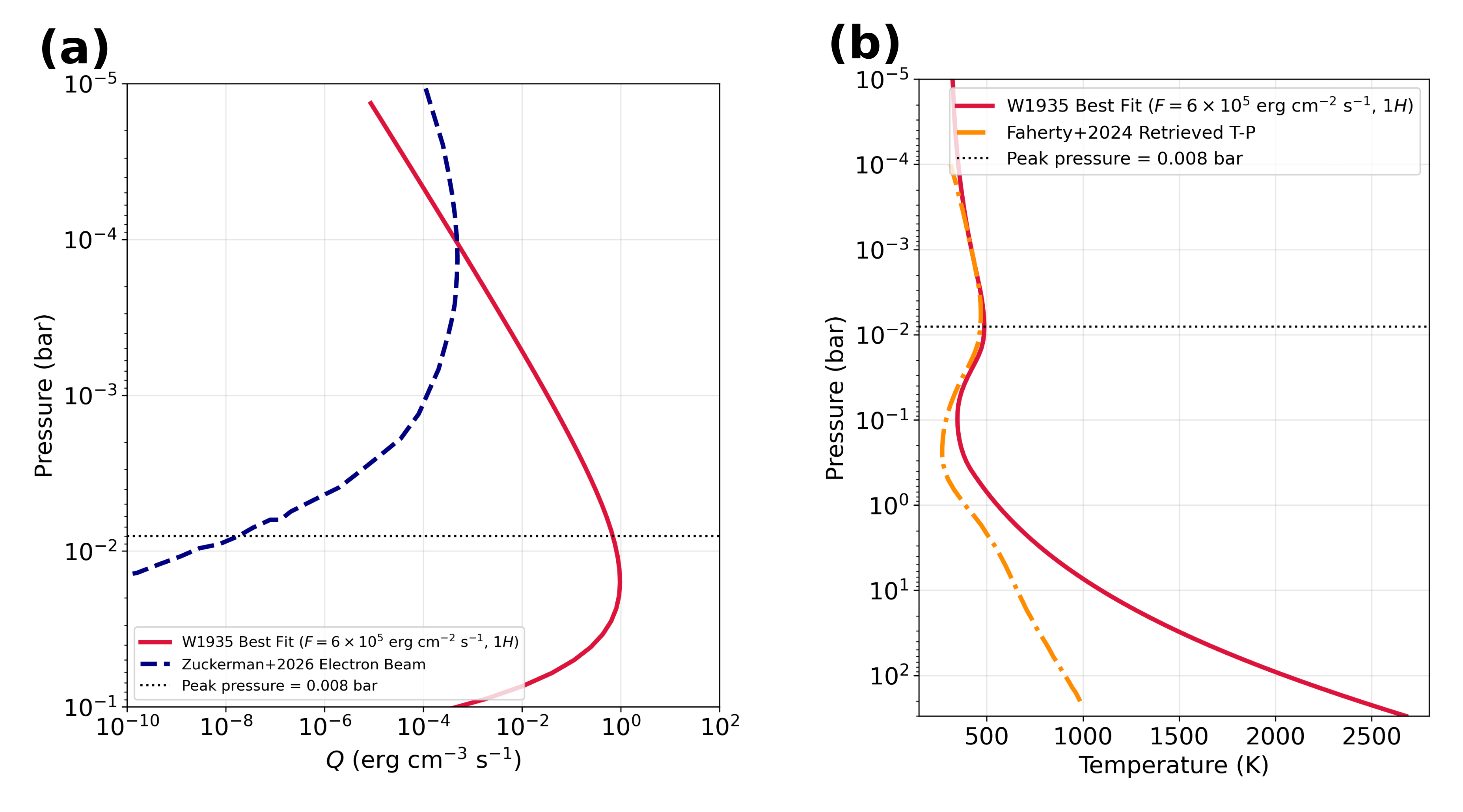}
    \caption{\textbf{(a)} Volumetric heating rate, $Q$, as a function of pressure. The red curve shows the heating profile inferred for the best-fitting W1935 model with an injected energy flux of $6\times10^{5}\,\mathrm{erg\ cm^{-2}\,s^{-1}}$ and $H/H_{\rm Chapman}=1.0$, while the blue dashed curve shows the electron-beam heating profile from \citet{2026ApJ...999...26Z}, converted from $\mathrm{eV\ m^{-3}\,s^{-1}}$ to $\mathrm{erg\ cm^{-3}\,s^{-1}}$. The horizontal dotted line marks the pressure of maximum energy deposition, $P_{\rm peak}=0.008$ bar. \textbf{(b)} Temperature--pressure profile of the best-fitting Chapman heating model for W1935 in red and the \cite{Faherty_2024} retrieved T-P profile in orange.}
    \label{fig:9}
\end{figure*}

As a benchmark for magnetically driven heating, we compare our inferred energy deposition rates to the electron-beam heating calculations of \citet{2026ApJ...999...26Z}. One of the models presented in that work considered a warmer brown dwarf atmosphere ($T_{\rm eff}\approx900$ K, $\log g \approx 5.0$), which we use as a point of comparison. For our best-fitting Chapman model of W1935 ($P_{\rm peak}=0.008$ bar, $H/H_{\rm Chapman}=1.0$), the required energy flux is $6\times10^{5}~{\rm erg~cm^{-2}~s^{-1}}$. Distributed over approximately one local atmospheric scale height, this corresponds to a characteristic volumetric heating rate of order $1~{\rm erg~cm^{-3}~s^{-1}}$ (Figure~\ref{fig:9}). By comparison, the electron-beam models of \citet{2026ApJ...999...26Z} span a range of heating rates depending on the assumed electron energy distribution, with their published example reaching peak volumetric heating rates of order $10^{14}~{\rm eV~m^{-3}~s^{-1}}$ ($\sim10^{-3}~{\rm erg~cm^{-3}~s^{-1}}$). Figure~\ref{fig:9} illustrates that, despite W1935 being approximately 500~K cooler than one of the brown dwarfs considered by \citet{2026ApJ...999...26Z}, substantially larger volumetric heating rates are required to reproduce its observed methane emission feature. Also, the energy injection is far too high in the atmosphere. Although higher-energy electron distributions can produce greater energy deposition, our best-fitting models require heating rates near the upper end of, or exceeding, those explored in current electron-beam calculations.

\begin{figure*}[htbp]  
    \centering
    \includegraphics[scale=0.7]{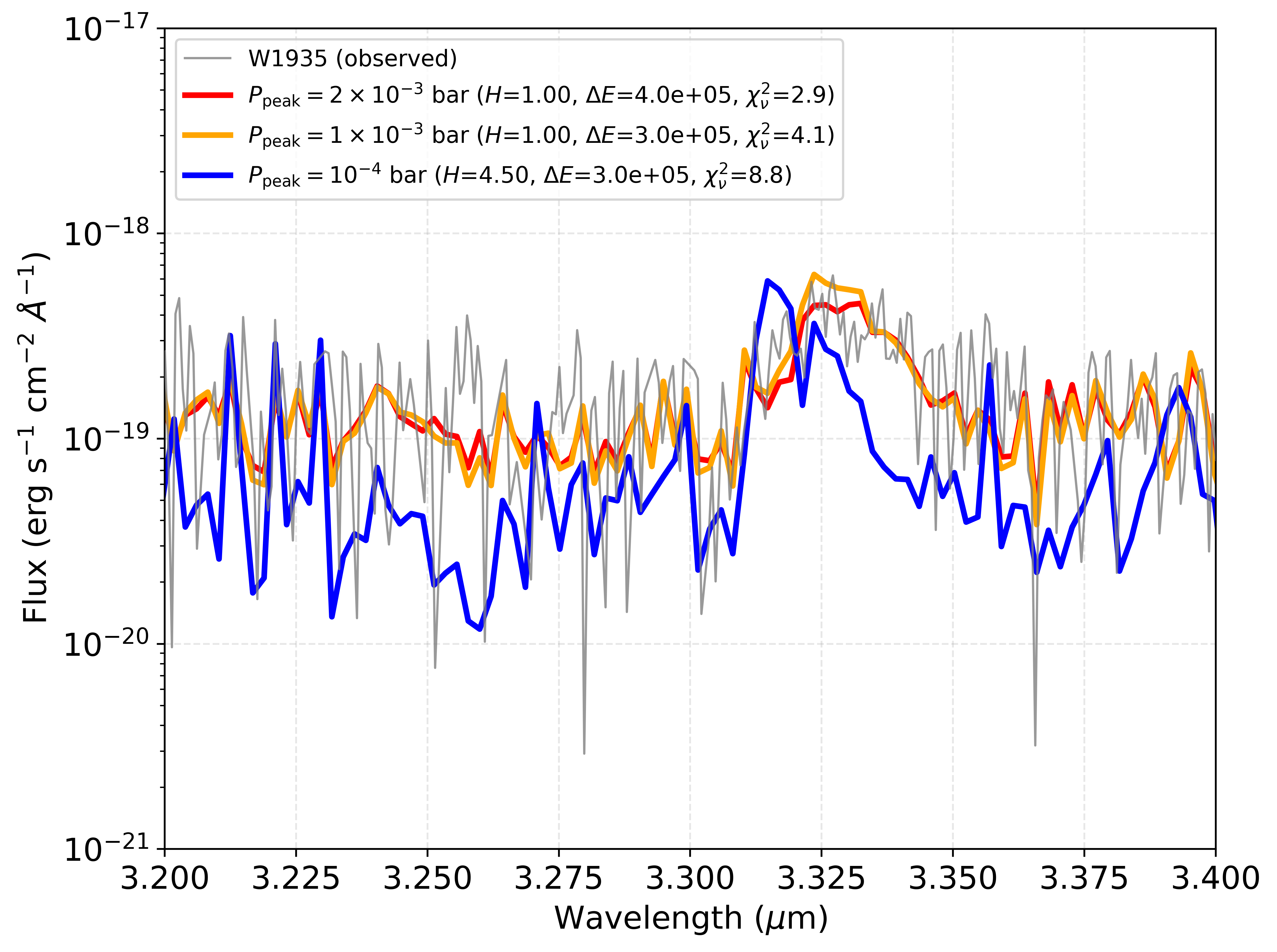}
    \caption {Comparison of representative Chapman heating models with progressively lower peak energy deposition pressures. The gray curve shows the observed W1935 spectrum, while the colored curves show the best-fitting models for $P_{\rm peak}=2\times10^{-3}$ bar (red), $10^{-3}$ bar (orange), and $10^{-4}$ bar (blue). All models produce methane emission near 3.3~$\mu$m, indicating that thermal inversions can form even when heating is deposited at very low pressures. However, as the peak deposition pressure decreases, the emission feature becomes increasingly narrow and less consistent with the observed spectral morphology. The broad methane emission feature observed in W1935 therefore favors energy deposition occurring near $10^{-2}$--$10^{-3}$ bar rather than exclusively in the uppermost atmosphere.}
    \label{fig:10}
\end{figure*}

This comparison should be interpreted with caution, as the electron-beam calculations were not tailored to W1935 and the resulting heating rates depend sensitively on the assumed precipitating electron population, magnetic field strength, and atmospheric structure. Additional exploratory Chapman models with peak deposition pressures of $10^{-3}$ and $10^{-4}$ bar also produce temperature inversions and methane emission, demonstrating that energy deposition at substantially lower pressures can still generate emission features. However, injecting energy at $10^{-4}$ bar or lower produce methane emission that is considerably narrower than the feature observed in W1935 (Figure~\ref{fig:10}). Thus, while the presence of a thermal inversion alone does not uniquely constrain the pressure of energy deposition, the shape and width of the methane emission feature favor heating distributed over deeper atmospheric layers near $P_{\rm peak}\sim10^{-2}$ and $\sim10^{-3}$~bar. These results suggest that the observed spectrum constrains not only the total energy deposited into the atmosphere, but also the vertical location over which that energy is released. For further comparison, at $P = 10^{-3}$~bar, Jupiter's stratospheric radiative heating rate is $\sim$80~erg~g$^{-1}$~s$^{-1}$ \citep{2001Icar..152..331Y}, corresponding to $\sim$30~erg~cm$^{-2}$~s$^{-1}$ in flux-divergence units (converted via $Q = H\rho H_{\rm scale}$, using Jupiter's atmospheric scale height and $g_{\rm Jupiter} \approx 2479$~cm~s$^{-2}$).

\section{Conclusions}
\label{sec:6}
W1935 is a brown dwarf discovered by the WISE mission and subsequently imaged by JWST. As an isolated, exceptionally cold object, the methane emission feature at 3.325 $\mu$m detected in its spectrum is particularly intriguing. Recent observations have confirmed that W1935 is a binary system \citep[]{defurio2025discoverysecondyydwarf}. \citet[]{Faherty_2024} demonstrated that reproducing this methane emission feature requires a temperature inversion in the upper atmosphere. This work quantifies the energy input required to generate a temperature inversion consistent with the observed methane emission feature using the the 1D radiative-convective climate models in \texttt{PICASO}.

\begin{enumerate}
    \item Section~\ref{sec:2.1} describes the Chapman energy injection mechanism implemented in \textit{PICASO}. Using a grid of deposition pressures and heating widths, we find that heating rates of order $10^{5}$--$10^{6}\ \mathrm{erg\ cm^{-2}\ s^{-1}}$ are sufficient to reproduce the observed thermal inversion and methane emission feature. The most successful models deposit energy at low pressures, indicating that upper-atmosphere heating plays a key role in shaping the thermal structure and emergent spectrum of W1935.
    \item Section~\ref{sec:3.1} examines the impact of disequilibrium chemistry relative to equilibrium models on the energy required to reproduce the methane emission feature in W1935. We find that, for identical injection depths and vertical extents, disequilibrium models require both a reduced local energy input and a lower total injected energy compared to equilibrium models.
    \item In Section~\ref{sec:3.2}, we predict the spectrum of W1935 over the $5$--$15\,\mu\mathrm{m}$ wavelength range to compare to future \textit{JWST} observations. Our models show that the methane emission feature observed at $3.325\,\mu\mathrm{m}$ also appears in the prominent $7.8\,\mu\mathrm{m}$ band, along with an ammonia emission feature near $\sim6\,\mu\mathrm{m}$. These spectral features are strongly sensitive to the magnitude of energy injection, the pressure level of deposition, and the vertical extent of the energy distribution, expressed in scale heights. Also since the original $T_{\rm eff}$ was computed assuming a normal brown dwarf spectrum outside of the then-available JWST data \citep{Faherty_2024} the `true' $T_{\rm eff}$ must be higher--by about $\sim15-50\,\rm K$ to account for this excess energy deposited needed to produce the inversion. 
    \item The resulting temperature--pressure profiles and $3$--$4\,\mu\mathrm{m}$ spectra for the best-fitting models are presented in Section~\ref{sec:3} for three representative Chapman scale height cases ($H/H_{\rm Chapman}=0.2$, 1.0, and 4.5). In Section~\ref{sec:3.3}, we account for spatially inhomogeneous heating through a fractional flux analysis to constrain the required energy deposition as a function of the emitting surface fraction. We find that as the emitting area increases, the localized energy input required to reproduce the observed spectrum correspondingly decreases.
    \item In Section~\ref{sec:4}, we explore two potential heating mechanisms, magnetically induced Joule heating (Section~\ref{sec:4.1}) and cometary bombardment (Section~\ref{sec:4.2}), to assess their ability to deposit sufficient energy into W1935's atmosphere and drive its temperature inversion. Joule heating is the more promising candidate, as it can deposit energy near the $\sim1-10$ mbar levels required by our models at field strengths orders of magnitude larger than Jupiter's, though it requires electron mixing ratios up to $\sim20$ orders of magnitude larger than predicted in chemical equilibrium and therefore an external ionization source such as auroral precipitation. Cometary bombardment can supply the total energy budget for sufficiently large or dense impactors, but it is unclear whether impacts would deposit that energy at the required pressures, and sustaining the heating would demand a substantial reservoir of cometary material and an efficient delivery mechanism. These results not only constrain the energetics of W1935's temperature inversion but also provide an important benchmark for interpreting atmospheric dynamics across the broader brown dwarf population, particularly for cold objects observed with \textit{JWST}. Notably we predict that the bolometric luminosity of W1935 will be found to be significantly higher when the radiated energy beyond $5\,\rm \mu m$ is accounted for.
    \item Comparison with the electron-beam heating models of \citet{2026ApJ...999...26Z} indicates that reproducing the observed thermal inversion in W1935 requires substantially greater energy deposition than predicted by one of the $T_{\rm eff}\approx900$ K brown dwarf model examined in that study. The best-fitting Chapman models require energy injection rates of order $10^{5}$--$10^{6}\,\mathrm{erg\,cm^{-2}\,s^{-1}}$, corresponding to characteristic volumetric heating rates of $\sim1$--$10\,\mathrm{erg\,cm^{-3}\,s^{-1}}$ near the region of peak energy deposition. These values exceed the ion- and electron-beam heating rates reported by \citet{2026ApJ...999...26Z} by several orders of magnitude, suggesting that significantly stronger particle precipitation or an alternative heating mechanism may be required to explain the thermal inversion in W1935 (Section~\ref{sec:5}).
    \item Gravity-wave heating remains a plausible explanation for the inversion, but confirming it will require dynamical models that capture both wave momentum and energy deposition. Such models also need to explain why so few Y dwarfs show this feature, perhaps wave deposition is sensitive enough to an object's specific properties that only rare cases like W1935 land in the right pressure range to produce an inversion. If so, the inversion may not be permanent, and could vary over time. Continued monitoring of the relevant emission features would help test both possibilities.
    \item The methane emission observed in W1935 requires strong upper-atmospheric heating, with inferred volumetric heating rates exceeding those explored in current electron-beam models \citep{2026ApJ...999...26Z}. While methane emission can be produced over a range of deposition pressures, heating at very low pressures ($\lesssim10^{-4}$ bar) generates emission features that are too narrow compared to the observations, favoring energy deposition near $10^{-3}$--$10^{-2}$ bar.
\end{enumerate}

Overall, our results demonstrate that reproducing the methane emission feature in W1935 requires a well-constrained energy input, and that this requirement is significantly reduced when disequilibrium chemistry is included. This highlights the critical role of chemical and radiative coupling in shaping the thermal structure of cold substellar atmospheres, where opacity redistribution can shift energy deposition and strengthen temperature inversions. More broadly, W1935 serves as a benchmark for understanding atmospheric heating processes and temperature inversions in isolated, cold brown dwarfs. As \textit{JWST} continues to observe similar objects, these results provide a framework for interpreting emission features and constraining the physical mechanisms responsible for energy injection in the coldest substellar atmospheres.

\begin{acknowledgements}
K.J.S. would like to thank her funding sources of this work: the University Fellows Program at the University of Arizona and the Carson Fellowship through the Lunar and Planetary Laboratory. J.M. acknowledges support from the National Science Foundation Graduate Research Fellowship Program under Grant No. DGE 2137420. The JWST data used in this article were obtained from the Mikulski Archive for Space Telescopes (MAST) at the Space Telescope Science Institute. The specific observations analyzed can be accessed via \dataset[doi: 10.17909/kcen-cj24]{https://doi.org/10.17909/kcen-cj24}.
\end{acknowledgements}

\bibliography{sample701}{}
\bibliographystyle{aasjournalv7}


\end{document}